\documentclass[twocolumn,amsmath,amssymb,letterpaper]{revtex4}

\usepackage{graphicx} \usepackage{dcolumn}

\begin{document}

\title{Spin-Based Quantum Computers made by Chemistry: Hows and Whys}

\author{Philip C. E. Stamp} \affiliation{Department of Physics and Astronomy,
and Pacific Institute of Theoretical Physics, University of British Columbia,
6224 Agricultural Road, Vancouver, Canada} \author{Alejandro Gaita-Ari{\~n}o}
\affiliation{Department of Physics and Astronomy, and Pacific Institute of
Theoretical Physics, University of British Columbia, 6224 Agricultural Road,
Vancouver, Canada}


\begin{abstract}

This introductory review discusses the main problems facing the
attempt to build quantum information processing systems (like
quantum computers) from spin-based qubits. We emphasize 'bottom-up'
attempts using methods from chemistry. The essentials of quantum
computing are explained, along with a description of the qubits and
their interactions in terms of physical spin qubits. The main
problem to be overcome in this whole field is decoherence - it must
be considered in any design for qubits. We give an overview of how
decoherence works, and then describe some of the practical ways to
suppress contributions to decoherence from spin bath and oscillator
bath environments, and from dipolar interactions. Dipolar
interactions create special problems of their own because of their
long range. Finally, taking into account the problems raised by
decoherence, by dipolar interactions, and by architectural
constraints, we discuss various strategies for making
chemistry-based spin qubits, using both magnetic molecules and
magnetic ions.

\end{abstract}

\maketitle

\section{Introduction}

The basic question addressed in this short review is: how might one
build a spin-based quantum computer using the 'bottom-up' methods of
chemistry? Before even starting on this topic, the reader might
suppose that we should first answer two other questions, viz. (i)
can one build a quantum computer at all? (ii) why would this be a
subject for chemists?

The answer to the first question is not known. Although one can
easily imagine various schemes for doing quantum information
processing, to actually build a 'Quantum Information Processing
System" (QIPS) one has to deal with a really quite fundamental
problem in Nature, usually called the 'decoherence problem'. Quantum
computing involves manipulating 'multiply entangled wave-functions',
involving many different sub-systems in the QIPS - in these terribly
complex quantum states, distributed over the whole QIPS, the
individual wave-functions of the individual components lose all
meaning and only the wave function of the entire system is
physically meaningful. The problem is that such states are extremely
delicate, and can be destroyed by very weak interactions with their
surroundings - and yet in order to use them we need to be able to
probe and manipulate them. Physicists and chemists working in
'nanoscience' have come up with many designs, and there has been
considerable success in building QIPS involving just a few entangled
'qubits' (the simplest kind of sub-system, involving only two
quantum states). This success has simply made the importance of the
decoherence problem more obvious - solving it will be the key that
unlocks the QIPS door (and doubtless many other doors as well).

Concerning the second question - so far the most successful designs
for QIPS have been made from ion traps\cite{ionT}, but it is not yet
clear how one might scale these up to the larger systems, involving
many entangled qubits, that are needed for quantum computation. The
most likely designs may be hybrids between 'solid-state' circuits of
qubits, with communication effected by photons - nobody really knows
yet. There are then 2 good reasons for concentrating on spin-based
systems, made using bottom-up methods. The first is that a device
which only uses spin degrees of freedom avoids moving charges around
- this is good because moving charges invariably causes decoherence.
The second is that if the basic sub-units are made by Nature (in the
form of molecules or ions) then quantum mechanics guarantees they
will all be identical, provided we can eliminate impurities,
defects, etc. This of course takes us into the realm of chemistry.

In writing this review we have assumed a reader who may not be too
familiar with this subject, but who is interested in understanding
the main ideas and problems, and who is also looking for suggestions
for future work in this area. We therefore begin, in section 2, by
describing briefly what a QIPS is, and how one would describe its
physical components if they were indeed made from spin qubits using
chemistry. This description also includes a discussion of the
physical mechanisms responsible for decoherence in such systems.
Given the importance of decoherence, we have devoted all of section
3 to how it works. The emphasis is on simple pictures, which often
convey a better feeling for what is going on than complex
mathematics. Then, in section 4, we outline some
features of the chemistry, and a few desiderata for any future
architecture. This allows us to make some suggestions
for future work. This last section is necessarily open-ended - at
present nobody knows which designs may ultimately succeed.

\section{Describing the system}

Physicists and chemists on the one hand, and computer scientists on
the other, have quite different ways of describing a system whose
ultimate aim is to do computing (quantum or classical). Each kind of
description is valuable, but the differences often cause
misunderstanding. In this section we describe and then connect the
two kinds of description; we then discuss an effective Hamiltonian
for a set of qubits made from real materials.

\subsection{Description in Computational Language}

\begin{figure}
\includegraphics[width=0.45\textwidth]{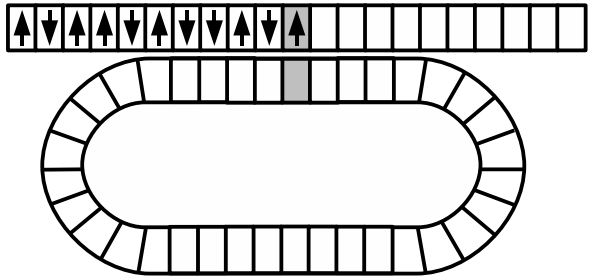}
\caption{Cartoon view of a Turing machine: a head (shown shaded) on a control loop reads and
writes onto a linear "tape". The states of both are binary (shown here as spins). We do not
show the register or the program parts of the machine.}
\label{Turing}
\end{figure}

Computer scientists like to reduce physical computational systems to
"Turing machines"\cite{turing}; a clear physical discussion of this
was given by Feynman\cite{feynC,feyn85}. In a Turing machine, a
"tape" moves forward or backwards past a "head"; the head has the
job of writing in or reading out particular discrete states
(labelled by different symbols) on the tape (and it may also erase
states). There is also a table of instructions which tell the tape
and head what to do (the "program"), and a "stable register" or
memory, which stores the state of the machine at any given time.
(Turing thought of this as being analogous to the state of mind of a
sentient being). The net result is shown in Fig.~\ref{Turing}. Note
that the state of a Turing machine at any time can be understood in
binary code as a string of 1's and 0's (or a sequence of on/off
states, or up/down, or heads/tails, etc.).

Even though real computers may not be constructed like a Turing
machine (indeed Turing machines provide a very clumsy model for a
realistic computing architecture), nevertheless any computer can be
mapped, in its logical and operational structure, to a Turing
machine. Moreover, Turing showed that a rather simply constructed
machine of this kind (since made even simpler) could do any discrete
computation. Thus all digital computers can be described in terms of
such "universal (classical) Turing machines".

This idea of classical Turing machines can be enlarged to cover
Quantum Turing machines, in which the notion of a state is
generalised to include the usual quantum superpositions of states,
in the head, tape, register and program\cite{feyn85,benioff,chuang}.
This generalisation seems straightforward - however the way one
describes such a system has led to much misunderstanding and a
severe divergence in the literature. Computer scientists like to
talk in terms of states, and logical operations on them - in quantum
computing these operations are called "unitaries". A "{\it quantum
computation}" is then a sequence of unitary transformations
$\hat{U}_N = \prod_k \hat{u}_k$ performed on some initial state
$|\Psi_{in}>$, to produce the final state $|\Psi_f>$. The initial
state is then the input to the computer, and the final state the
output.

These states are a quantum mechanical generalisation of the binary
strings mentioned above - however they are much more complicated.
Consider, e.g., a classical pair of binaries, which can be in one of
four states (ie, (11), (10), (01), or (00)). In quantum mechanics,
systems that have 2 states are called 'two-level systems' (TLS), or
'qubits'. Now however a single qubit can exist in an arbitrary
superposition of states $\vert \Psi_1 \rangle = a_0 e^{i\phi_0}
\vert 0 \rangle + a_1 e^{i\phi_1} \vert 1 \rangle$, where the
coefficients $a_0, a_1$ are real, $a_0^2 + a_1^2 = 1$, the states
$\vert 0 \rangle$ and $\vert 1 \rangle$ correspond to the original
classical states (0) and (1), and $\phi_0, \phi_1$ are phases. A
{\it pair} of qubits has the general wave-function
\begin{eqnarray}
 \vert \Psi_2 \rangle &=& a_{00}e^{i\phi_{00}}
\vert 00 \rangle + a_{10}e^{i\phi_{10}} \vert 10 \rangle \nonumber \\ &+&
a_{01}e^{i\phi_{01}} \vert 01 \rangle + a_{11} e^{i\phi_{11}} \vert 11 \rangle
 \label{2QB}
\end{eqnarray}
and we see that if we have $N$ qubits then we will end up with $2^N$
coefficients and $2^N$ 'relative phases' (of which one is
redundant). The great power of quantum computation resides in all
the information that can be stored in these phases, allowing one to
'superpose' many computations at once\cite{chuang}. A key feature of
these states is that typically the $N$ qubits are '{\it entangled}';
one cannot write them as a product over independent wave-functions
for each individual qubit. As first noted by Einstein\cite{EPR35},
this leads to very counter-intuitive features - it is simply not
correct to assign a quantum state to an individual qubit, and only
the $N$-qubit wave-function, existing with all its relative phases
in a vast $2^N$-dimensional Hilbert space, is physically meaningful.

\begin{figure}
\includegraphics[width=0.40\textwidth]{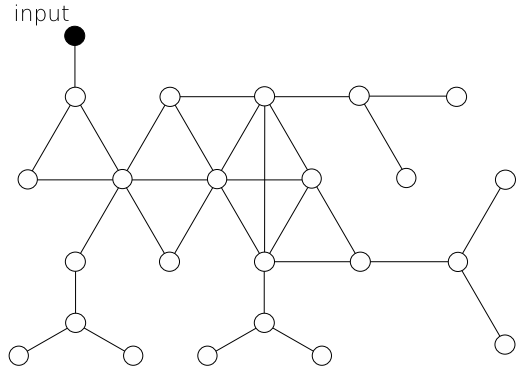}
\caption{In a quantum walk, the system "hops" between nodes, each of which
represents a state in the Hilbert space of the QUIPS. Allowed transitions are
shown as links between the nodes.}
\label{QWfig}
\end{figure}

The above description is very abstract. To get a more intuitive
understanding, it is helpful to map the QIPS onto a "quantum walk"
system, in which a single particle hops around a graph. Each node of
the graph represents a state in the Hilbert space of the QIPS, and
the links between the nodes represent the transition matrix elements
(in general time-dependent) between them. Thus the time evolution of
a QIPS can be represented as a "quantum walk" in information
space\cite{QW}; and it is always possible to make this mapping for a
QIPS composed of qubits.

To see how this works one needs to specify a Hamiltonian, to give
the system dynamics. The simplest quantum walk Hamiltonians take the
form\cite{Hines07}:
\begin{equation}
 \hat{H}_{QW}=\sum_{\mu\nu}\left[T_{\mu\nu}(t)c^{\dagger}_\mu
c_\nu + H.c. \right]+\sum_{\mu} \epsilon_\mu(t)c^{\dagger}_\mu c_\mu
 \label{QW}
\end{equation}
where the operators $c^{\dagger}_\mu$, $c_\mu$, create or destroy a
particle at the $\mu$-th node, $\epsilon_\mu(t)$ is the $\mu$-th
node energy, and $T_{\mu\nu}(t)$ the non-diagonal hopping matrix
element between nodes $\mu$ and $\nu$.  This Hamiltonian must be
generalized to describe "composite quantum walks" (where discrete
internal "spin" variables at each node can also couple to the
walker) if we are to fully describe a QIPS\cite{Hines07}. The great
advantage of this representation, at least for physicists and
chemists, is that we have a good intuitive understanding of how
particles move around such graphs. One also clearly sees where
things can get complicated, if the $\epsilon_\mu$ and $T_{\mu\nu}$
are strongly disordered or if their time-dependence is non-trivial.
How this works in practice can be found in the
literature\cite{QW,Hines07}.

\subsection{Physical Description of a QIPS}

We now turn to a description of a QIPS of the kind favoured by
physicists and chemists, one which will allow them to design and
make one, written in terms of the actual physical components of the
system. So let us now start from a much more physical model, using a
Hamiltonian written in terms of the qubits themselves:
\begin{equation}
 \hat{H}_{QB} = \sum_j\vec{B}_j(t)\cdot\hat{\tau}_j
+
\sum_{ij}U_{ij}^{\alpha\beta}(t)\hat{\tau}_i^\alpha\hat{\tau}_j^\beta
 \label{QB}
\end{equation}
Here the Pauli operator $\hat{\tau}_j$ operates on the j-th qubit,
and both the local fields $\vec{B}_j(t)$ and the interaction
$U_{ij}^{\alpha\beta}(t)$ can be time-dependent.

Now qubits are real physical objects, and many different physical
systems have been proposed to make them. Solid-state examples which
have achieved some experimental success include spins in
semiconductors and quantum dots\cite{QDot}, various designs based on
superconductors\cite{SCQB}, vacancy centers in diamond\cite{NV},
single molecule magnets (SMMs)\cite{SMMQB}, and rare earth
spins\cite{REQB}. To focus the discussion here, let us consider an
example.

\begin{figure}
\includegraphics[width=0.35\textwidth]{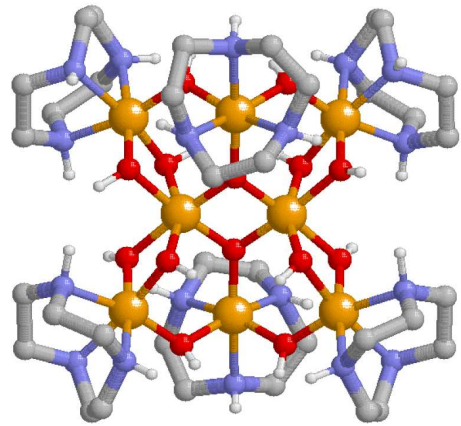}
\caption{Simplified ball-and-stick representation of
[Fe$_8$O$_2$(OH)$_{12}$(tacn)$_6$]$^{8+}$ single molecule magnet,
showing the octahedral iron ions, the oxo and hydroxo bridges and
the 1,4,7-triazacyclononane (tacn) ligands completing the
coordination spheres.} \label{Fe8}
\end{figure}

{\it An SMM Example}: A "Single Molecule Magnet" (SMM) typically has
a magnetic core of transition metal (TM) ions, surrounded by a
protective cage of organic ligands; the inter-ion couplings are
usually antiferromagnetic.  A useful (because well-understood)
example is the "Fe$_8$" molecule\cite{Fe8mol}, shown in
Fig.\ref{Fe8}, in which a core of eight Fe$^{3+}$, spin-$5/2$ ions
forms a total spin-10 system.

The simplest (and very commonly employed) model of a system like Fe$_8$ starts
by reducing the spin Hamiltonian
\begin{equation}
 \hat{H}_{SS} =
\sum_{ij}J_{ij}^{\alpha\beta}\hat{s}_i^\alpha\hat{s}_j^\beta +
\sum_{j}k_{j}^{\alpha\beta}\hat{s}_j^\alpha\hat{s}_j^\beta
 \label{SS}
\end{equation}
written in terms of the individual TM spins (for simplicity we
ignore here terms like Dzyaloshinski-Moriya interactions) to a single "giant
spin" model of form
\begin{equation}
 \hat{H}_{GS} =
K_{2}^{\alpha\beta}\hat{S}_\alpha\hat{S}_\beta +
K_{4}^{\alpha\beta\gamma\delta}\hat{S}_\alpha\hat{S}_\beta\hat{S}_\gamma\hat{S}_\delta
+ \ldots
 \label{GS}
\end{equation}
where we assume all low-energy states have
the same total spin $\vec{S}=\sum_j\vec{s}_j$. It is very common in the
literature to drop many of the higher terms in this - for example, in the case
of Fe$_8$ one usually writes the simple quadratic biaxial form
\begin{equation}
 {\cal H}({\bf S}) = -DS^2_z + E S^2_x + K^{\perp}_4 (S^4_{+}+S^4_{-}) - \gamma
{\bf S} \cdot {\bf H}_{\perp},
 \label{HS10}
 \end{equation}
which gives a '2-well' crystal field potential acting on $\vec{S}$.
Here $D/k_{\rm B} = 0.23$ K, $E/k_{\rm B} = 0.094$ K,
$K^{\perp}_4/k_{\rm B} = -3.28 \times 10^{-5}$ K, $\gamma = g_e
\mu_B \mu_0$, and $g_e \approx 2$ is the isotropic $g$-factor of the
spin-10 moment. All the higher terms $\sim O(S^6)$ in (\ref{GS}) are
thus dropped. This effective Hamiltonian for Fe$_8$ is used below
roughly $5-10K$; above this one certainly has to worry about other
states, contained in the more general 10-spin Hamiltonian
(\ref{SS}).

\begin{figure}
\includegraphics[width=0.45\textwidth]{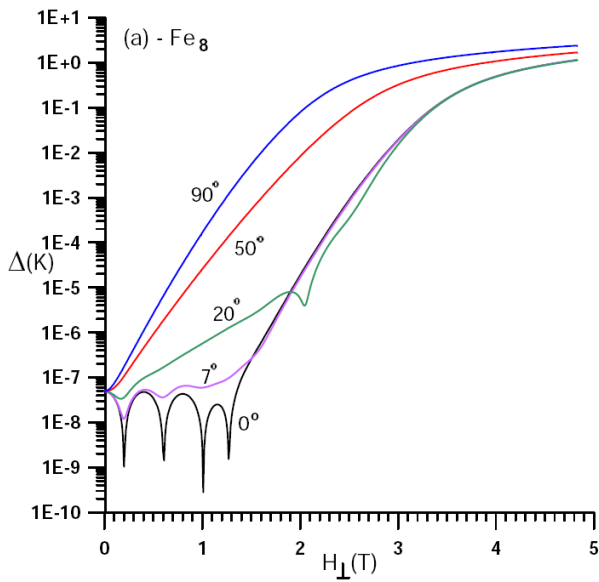}
\caption{Variation of the effective transverse field $\Delta_o$ by
application a real transverse field $\vec{H}_0^{\perp}$ at different
angles in the Fe$_8$ system.} \label{graphFe8}
\end{figure}

At still lower energies one can truncate (\ref{GS}) to its lowest singlet or
doublet - in the latter case we have a qubit Hamiltonian of form
\begin{equation} \hat{H}^0_{QB}=\vec{B}(\vec{H}_0)\cdot\vec{\tau}
 \label{QB0}
\end{equation}
where the effective field $\vec{B}(\vec{H}_0)$ depends both in the
crystal field potential and the applied field $\vec{H}_0$. A very
important special case of this arises for the 2-well potential,
where there is a 'natural basis' for the states, in which $\vert
\uparrow \rangle$ represents the state localised in one well and
$\vert \downarrow \rangle$ the state localised in the other. One
then often writes
\begin{equation}
\hat{H}^0_{QB}\rightarrow\Delta_0(\vec{H}_0^{\perp})\hat{\tau}_x +
\epsilon_0(\vec{H}_0^{\parallel})\hat{\tau}_z
 \label{QB0arrow}
\end{equation}
so that $B_x(\vec{H}_0)=\Delta_0(\vec{H}_0^{\perp})$; here
$\vec{H}_0^{\perp}$, $\vec{H}_0^{\parallel}$ are the applied fields
perpendicular and parallel to the Fe$_8$ easy axis. We can think of
$\Delta_o$ as an "effective transverse field", driving transitions
between the $\vert \uparrow \rangle$ and $\vert \downarrow \rangle$
states of the qubit. This effective field can be varied by applying
the real external transverse field $\vec{H}_0^{\perp}$ - how it does
this is shown for Fe$_8$ in Fig~\ref{graphFe8}. The 2-well system
possesses a real advantage - it is protected against external
transverse field fluctuations, which have a small effect on
$\Delta_o$.

\vspace{2mm}

\begin{figure}
\includegraphics[width=0.45\textwidth]{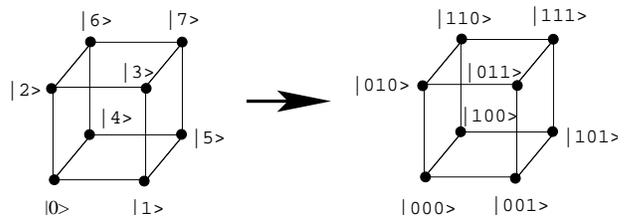}
\caption{The states of a 3-qubit system, with 8 states, shown at
left in quantum walk representation, and at right in the qubit
representation.}
 \label{Qbit-lattice}
\end{figure}

{\bf Connection between Quantum Walk and Qubit formulations}: We
have seen that a quantum computer can be thought of as a particle
'walking' in information space, described by a Hamiltonian like
(\ref{QW}), or as a set of interaction qubits with Hamiltonian
(\ref{QB}). So how do we connect these two? The answer is shown in
Fig.~\ref{Qbit-lattice} for a set of 3 interacting qubits - in
general the $2^N$ states of an $N$-qubit system map to the corners
sites of an $N$-dimensional hypercube, and the quantum walker hops
from one site to another on this hypercube.

How it hops depends on the exact details of the qubit Hamiltonian.
Although it is too lengthy to give here, the explicit mapping
between the quantum walk Hamiltonian $\hat{H}_{QB}$ in (\ref{QB})
and the qubit Hamiltonian $\hat{H}_{QW}$ in (\ref{QW}), for general
qubit couplings, is fairly straightforward - it was given in
ref.\cite{Hines07}. Another commonly used representation is in terms
of "quantum gates"\cite{chuang}, which can also be mapped to quantum
walks\cite{Hines07}, and thus also allows a connection to the
quantum Turing machine language.

We have seen above how to describe a QIPS in terms of interacting
qubits, of a quantum walker, and of a quantum Turing machine (and
there are other descriptions - for example, we have for reasons of
space ignored adiabatic quantum computation entirely). But how well
do such toy models match the real world in which physicists and
chemists work? We now address this question.

\subsection{Realistic Models of QIPS}

Obviously both the general models (\ref{QW}), (\ref{QB}), as well as
the specific models (\ref{SS}), (\ref{GS}), and (\ref{QB0arrow}) for
magnetic molecules, are oversimplified. Many papers have appeared
discussing both the failures of general models like (\ref{QW}),
(\ref{QB}), or of the simple qubit forms (\ref{QB0}) and
(\ref{QB0arrow}), as well as more specific corrections to the giant
spin model (\ref{GS}). Let us begin by enumerating some of the
problems of such models as applied to SMM's, and then discuss the
more general case. The principal shortcomings are as follows:
\begin{enumerate}
\item The giant spin model does
not include many other discrete electronic degrees of freedom (spin
and charge); if these lie at low enough energies, they couple to
$\vec{\tau}$ (and of course to $\vec{S}$). Because they are
discrete, they behave like "loose spins" coupling to $\vec{\tau}$
(see refs.\cite{PS93,PS96}). This happens, for example, if one has
structural transformations in the SMM (e.g. Jahn-Teller
distortions), or if some exchange couplings $J_{ij}$ are much weaker
than others. Note that in a multi-SMM system, there will also be
impurities, and "rogue molecules" which are different from the usual
SMM because of some internal distorsion, or because of stray local
fields created by, e.g., dislocations. These will also behave like
"loose spin" degrees of freedom coupling to $\hat{\tau}$.
\item In reality $\vec{\tau}$ will couple to
many other 'environmental modes' in the vicinity of the SMM. This
includes any nuclear spins in the SMM, coupled via hyperfine
interactions, nuclear spins outside the SMM, coupled via magnetic
dipolar or possibly transfer hyperfine interactions\cite{PS93,PS96}.
Also $\vec{\tau}$ will couple to photons (via magnetic dipole and
electric dipole couplings) and both extended phonon modes (i.e.,
magnetoacoustic couplings\cite{PS96,villain96}) and local phonons
(e.g., vibrons, or impurity-induced local phonons). In a conducting
system, $\vec{\tau}$ can also couple to extended electronic
excitations in various ways\cite{PS96}.
\item Finally, an ensemble of SMMs, as in a QIPS, will have
various 'stray' couplings between the $\{\vec{\tau}_j\}$, mediated
by all the degrees of freedom discussed above, and not included in
the 'bare' interactions $U_{ij}^{\alpha\beta}(t)$ which already
appear in (\ref{QB}) above. These include 'slow' couplings like
those mediated by phonons, in which retardation effects can be
important.
\end{enumerate}

Actually there is nothing specific to SMM's about problems like
these. Quite generally, in describing a QIPS, we need to include (i)
coupling of the qubits to stray environmental degrees of freedom,
and (ii) stray couplings between the qubits from these environmental
degrees of freedom.

So what kind of model should we really be using for a set of coupled
SMM qubits, instead of (\ref{QB})? The answer proceeds in 2
stages.

\vspace{2mm}

{\bf (i) Environmental Modes}: We can divide all the environmental
modes into 2 classes\cite{PS00}. First, for all the discrete {\it
localized} modes, like defects, nuclear spins, "loose" spins,
localized phonons, etc, we can use a model which couples the qubits
to a "spin bath" environment\cite{PS00} of localized modes. In its
simplest form this describes a set $\{\vec{\sigma_k}\}$ of two-level
systems, and we can generalize the bare qubit Hamiltonian in
(\ref{QB}) to
\begin{equation}
 \hat{H}_{QB} \;=\;
H^0_{QB}(\vec{\tau})+
H^0_{SB}(\vec{\sigma_k})+H^{SB}_{int}(\tau_i,\{\vec{\sigma}_k\})
 \label{H-SB}
\end{equation}
in which the Pauli matrices $\{\vec{\sigma}_k\}$ act on the
2-level systems, and where the spin bath has its own Hamiltonian
\begin{equation}
 \hat{H}_{SB}^0(\{\sigma_k\})=\sum_k
\vec{h}_k\cdot\vec{\sigma}_k+
\sum_{kk'}V_{kk'}^{\alpha\beta}\sigma_k^{\alpha}\sigma_{k'}^{\beta}
 \label{H-SBo}
\end{equation}
in which the local fields $\{\vec{h}_k\}$
typically dominate over the rather weak interspin coupling
$V_{kk'}^{\alpha\beta}$; and finally the qubit couples to the spin bath via the
term
\begin{equation}
 H^{SB}_{int}(\tau_i,\{\vec{\sigma}_k\})=\sum_k
\omega_k^{\alpha\beta}\tau^{\alpha}\sigma_{k}^{\beta}
 \label{H-SBint}
\end{equation}
with couplings $\omega_k^{\alpha\beta}$. More generally we have to
couple to a spin bath in which the individual modes have more than 2
levels -- for example, to a set of nuclear spins with spin $I$,
having $2I+1$ levels each. This description may look rather
complicated, but note that we can also write $\hat{H}_{QB}$ in the
more transparent form
\begin{equation}
 \hat{H}_{QB}
\;=\;H^0_{QB}(\vec{\tau})+ \sum_k \left( \vec{\gamma}_k +
\vec{\xi}_k \right) \cdot \vec{\sigma}_k
 \label{H-QB-red}
 \end{equation} where we define the time-dependent vectors
$\vec{\gamma}_k(t)$ and $\vec{\xi}_k(t)$, having components
\begin{eqnarray}
\gamma_k^{\alpha}(t) &=& h_k^{\alpha} +
\sum_\beta\omega_k^{\alpha\beta}\tau_{\beta}(t) \nonumber \\
{\xi}_k^{\alpha}(t) &=& \sum_{k'}\sum_\beta
V_{kk'}^{\alpha\beta}\sigma_{k'}^{\beta}(t)
 \label{gam-xi}
\end{eqnarray}
Equation (\ref{H-QB-red}) simply says that the $k$-th bath spin
moves in a dynamic field coming from both the qubit (incorporated in
$\vec{\gamma}_k$(t)) and the other bath spins (incorporated in
$\vec{\xi}_k$(t)). Note that $\vec{\gamma}_k$(t) is the sum of a
static field $\vec{h}_k$ and a dynamic field coming from the qubit.
Of course the motion of the $k$-th bath spin (and all the other bath
spins) will also react back on the qubit via the interaction
$\omega_k^{\beta\alpha}$.

\begin{figure}
\includegraphics[width=0.50\textwidth]{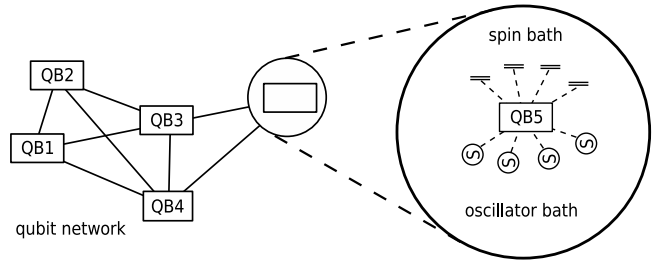}
\caption{A network of qubits interacts via interactions $J_y(t)$.
However each qubit is also interactiong with a "spin bath" and an
"oscillator bath" (for details see text).}
\label{fig5}
\end{figure}

As an example consider the Fe$_8$ molecule again, in which the core
of eight Fe$^{3+}$ spin-5/2 ions couples to over 200 nuclear spins
in each molecule (these being a mixture of $^1$H, $^2$H, and
possibly $^{17}$O and $^{57}$Fe spin 1/2 nuclear spins, as well as
some higher spin Br and N nuclear spins). In this system, analysed
in detail in refs\cite{TupS04,TupS08}, the
$\{V_{kk'}^{\alpha\beta}\}$ are weak internuclear couplings
($|V_{kk'}^{\alpha\beta}|<10^{-7}K$ even for nearby nuclei), the
$\{\omega_k\}$ are hyperfine couplings (with
$|\omega_k|\simeq$3-4$mK$ for $^{57}$Fe nuclei or nearby proton
spins) and the $\{\vec{h}_k\}$ include both the weak nuclear Zeeman
couplings, and a term arising when the Fe$_8$ giant spin is not
flipping between antiparallel orientations when it tunnels.

What about {\it delocalized} modes, like phonons or photons? These
can be handled in a standard way, by coupling the qubit to a bath of
oscillators\cite{AJL87}, which now adds an oscillator term
\begin{equation}
\hat{H}_{OB}\left(\{x_q\}\right)= {1 \over 2}\sum_q\left(\frac{p_q^2}{m_q}+
m_q\omega_q^2x_q^2\right)
 \label{H-OB}
\end{equation}
plus a coupling
\begin{equation}
 \hat{H}^{OB}_{int}\left(\vec{\tau},\{x_q\}\right)= \sum_q
c_q^{\alpha}\tau^{\alpha} x_q
 \label{H-OBint}
\end{equation}
to the qubit. In a SMM system, $x_q$ would be e.g. the coordinate of
a phonon with momentum $\vec{q}$, $\omega_q$ its energy, and
$c_q^{\alpha}$ is related to the magnetoacoustic
coupling\cite{deLach} between the SMM and this phonon. This coupling
takes the form
\begin{equation}
 V_{s\phi} = \sum_q g_{\vec{q}}
\hat{O}_S(\vec{S}) \hat{O}_{\phi}(x_q)
 \label{Vsphi}
\end{equation}
where $\hat{O}_S(\vec{S})$ is an operator acting on the giant spin
$\vec{S}$, $\hat{O}_{\phi}(x_q)$ an operator acting on the phonons,
and $g_{\vec{q}}$ a coupling function; typically $\vert g_{\vec{q}}
\vert \sim 10-100~K$ for a SMM. When we then truncate the giant spin
Hilbert space down to the qubit space, we get a coupling of the form
in (\ref{H-OBint}).

{\bf Stray Interactions}: Now consider the interaction
$U_{ij}^{\alpha\beta}(t)$ between the qubits. Essentially this
incorporates all interactions mediated by fields which operate so
fast that they can be treated as instantaneous on the timescale of
the QIPS. This includes all photon-mediated interactions (including
dipolar interactions and exchange interactions). However it does
{\it not} include phonon-mediated interactions, which are slow - if
the time $t_c \sim c_s/L$ taken for phonons to travel across a QIP
of size $L$ at velocity $c_s$ is not much less than the timescale of
switching or operation of qubits, then we must treat these
interactions as retarded, and this can actually lead to decoherence.
We ignore this problem in what follows but it is important. This
spin-phonon interaction typically has a complicated form - however
the main part will be of dipolar form, of strength $\sim g_o^2/\rho
c_s^2 r_{ij}^3$, where $g_o$ is the $\vec{q} = 0$ component of
$g_{\vec{q}}$ above.

It is essential to distinguish between the short-range and
long-range parts of $U_{ij}^{\alpha\beta}(t)$; we write
\begin{equation} U_{ij}^{\alpha\beta}(t)
\;=\; K_{ij}^{\alpha\beta} + D_{ij}^{\alpha\beta}
 \label{U-ij}
 \end{equation}
where $K_{ij}$ is the short-range interaction (typically exchange or
superexchange), with strength $\sim 50-1000~K$ for transition metal
spins and $\sim 0.1-5~K$ for rare earths; and $D_{ij}$ is the
magnetic dipolar interaction, of strength $\sim V_D/r_{ij}^3$, where
$V_D \sim (Sa_o)^2~K$, where $a_o \sim 1{\AA}$ is a microscopic
lattice length. The key point is that $K_{ij}$ is only appreciable
for nearest-neighbour systems - it falls off exponentially and very
fast with distance.

\section{Decoherence}

Decoherence is \underline{the} fundamental problem blocking the
manufacture of any QIPS. It is considered by many to be an
insuperable problem (an opinion we do not share!), as well as being
at the very heart of our understanding of quantum
mechanics\cite{decoherence}.

For anyone who wants to make a QIPS, decoherence creates a very
practical problem - since a QIPS will not work if there is too much
decoherence, then in designing one, we need to know how big the
decoherence will be. This question has a chequered history. Early
calculations of decoherence were far too optimistic - they used
oscillator bath models, but at low temperatures most decoherence
comes from spin bath environments of localised modes.

Our purpose in this section is (i) to give readers an intuitive
feeling for how decoherence works in practise in a system of qubits,
and (i) give some results for decoherence rates which we hope will
help to better design a QIPS.

Quite generally, decoherence in the dynamics of some system A is
caused by its entanglement with its environment\cite{decoherence}. A
key result of quantum mechanics says that if we do not
\underline{know} what the environment is doing, then we must average
over all of its possible states when calculating the dynamics of A.
But if the environment is entangled with A, then this will "smear
out" the phase dynamics of A. Another way of putting this is that if
the the environment is reacting to the dynamics of A, then it is
actually in effect measuring it - and this actually destroys phase
coherence in this dynamics.

To see how decoherence works, it is best to divide the discussion
between the 3 main mechanisms.

\subsection{Spin bath decoherence}

\begin{figure}
\includegraphics[width=0.35\textwidth]{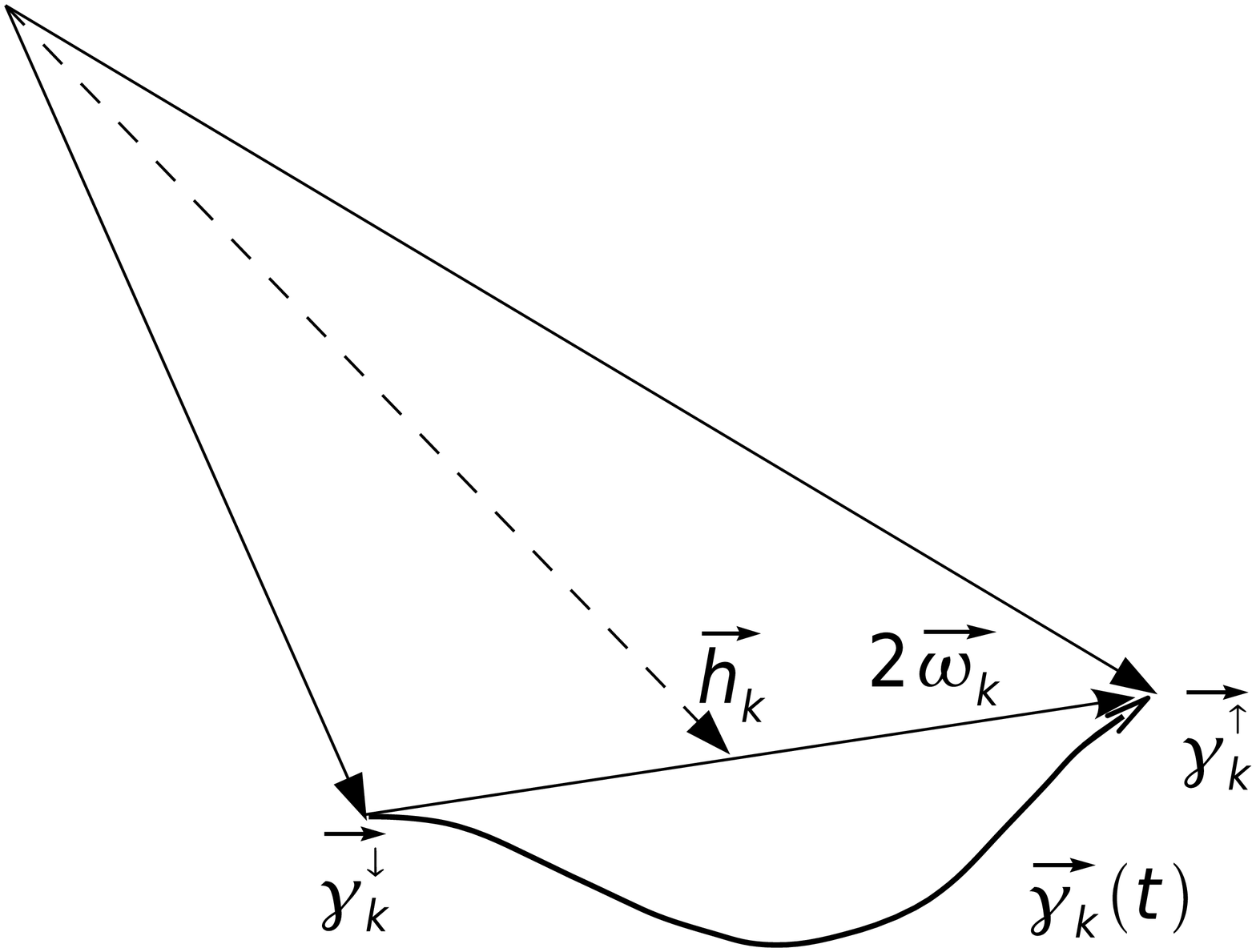}
\caption{The k-th bath spin coupled to a qubit has a field
$\vec{\gamma}_k(t)$ acting on it, a sum of a static field
$\vec{h_k}$, and a part which varies rapidly from $-\vec{\omega_k}$
to $\vec{\omega_k}$ when $\vec{\gamma}_k(t)$ moves between
$\vec{\gamma}_k^{\downarrow}$ and $\vec{\gamma}_k^{\uparrow}$; this
happens when the qubit flips from $|\downarrow>$ to $|\uparrow>$ (a
typical path for $\vec{\gamma}_k(t)$ during this flip is shown). }
\label{gammaF}
\end{figure}
To understand how spin bath decoherence works we need first to see
how the bath spins move in the presence of the qubit. Recall that
the $k$-th bath spin moves in a dynamic field $\vec{\gamma}_k(t) +
\xi_k(t)$ (see equations (\ref{H-QB-red}) and (\ref{gam-xi})). Since
$\xi_k$ is very small compared to $\vec{\gamma}_k$, let us begin
with $\vec{\gamma}_k$, the net field on the k-th bath spin
$\sigma_k$; recall it is the sum of the static field $\vec{h}_k$ and
the time-varying field from the qubit. Now $\vec{\gamma}_k$ spends
almost all of its time in the quiescent states
$\vec{\gamma}_k^{\uparrow}$ and $\vec{\gamma}_k^{\downarrow}$,
associated with the qubit states $|\uparrow>$ and $|\downarrow>$,
and given by
\begin{equation}
 \gamma_k^{\sigma}=\vec{h}_{k}+\sigma
\vec{\omega}_k
 \label{ksigma}
\end{equation}
where $\sigma=\pm$ corresponds to $|\uparrow>$, $|\downarrow>$ and
the vector $\vec{\omega}_k$ is defined by its components
$\omega_k^\alpha\equiv\omega_k^{z\alpha}$ (see Fig.~\ref{gammaF}).
However every time the qubit flips (taking a very short "bounce
time" $\tau_B \equiv 1/\Omega_o$ to do so), the vector
$\vec{\gamma}_k$ moves rapidly between these two quiescent states.

The vector $\vec{\xi}_k$ defining the field from the other bath
spins also varies in time, but quite differently. Typically
$\vec{\xi}_k$ will look much like a small and slowly-varying random
noise source, adding a small "jittering" perturbation on
$\vec{\gamma}_k$, to give the total time-dependent field acting on
$\vec{\sigma}_k$.

\begin{figure}
\includegraphics[width=0.35\textwidth]{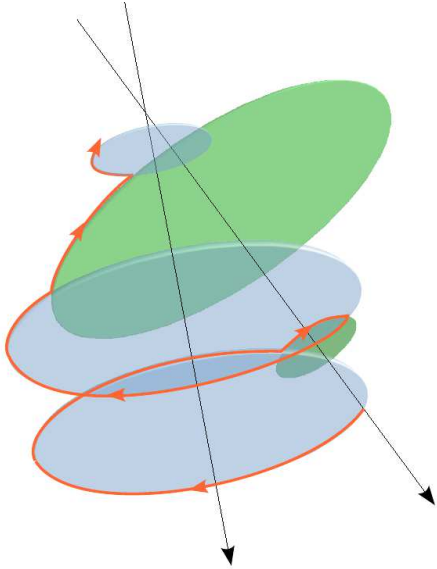}
\caption{ A spin qubit flipping between its up and down quantum
states also flips the field acting on nearby nuclear spins between
two orientations (shown as arrows). The nuclear spins try to precess
in this qubit field, but each time it suddenly changes they must
begin anew. Thus, the path a nuclear spin vector follows (an example
is shown in red) is conditional on the specific trajectory of the
qubit, ie., the two are quantum-mechanically entangled.  }
\label{precessF}
\end{figure}

How does $\vec{\sigma}_k$ itself move in response to this, as
$\vec{\gamma}_k$ "jerks" back and forth between
$\vec{\gamma}_k^{\uparrow}$ and $\vec{\gamma}_k^{\downarrow}$? The
answer is shown in Fig~\ref{precessF}. At $t=0$, $\vec{\sigma}_k$
will begin to precess about the initial field $\gamma_k(t=0)$ (which
is, say, $\vec{\gamma}_k^{\uparrow}$). However as soon as
$\vec{\gamma}_k^{\uparrow}$ jumps to $\vec{\gamma}_k^{\downarrow}$,
then $\vec{\sigma}_k(t)$ must begin to precess around the new field.
The resulting motion of $\vec{\sigma}_k$ depends on the exact time
sequence of flips of $\vec{\gamma}_k$.

Actually there are two small corrections to this picture. First, there is a
small extra "wobble" caused by $\vec{\xi}_k(t)$. Second, we have described the
motion of $\vec{\gamma}_k^{\downarrow}$ as though the jumps between
$\vec{\gamma}_k^{\uparrow}$ and $\vec{\gamma}_k^{\downarrow}$ were
instantaneous, so that $\vec{\sigma}_k$ cannot follow them. However this is not
true -- although they are sudden, there is still a small perturbation on the
dynamics of $\vec{\sigma}_k$.

Now notice a crucial consequence of this picture - it is that the path followed
by $\vec{\sigma}_k$ is {\it conditional} on that followed by $\vec{\gamma}_k$.
This means that the 2 systems are entangled - indeed rather strongly so, since
the path followed by $\vec{\sigma}_k$ depends more or less entirely on the time
evolution of $\vec{\gamma}_k$ (apart from the small perturbation coming from
$\vec{\xi}_k$).

Now we can see how a spin bath will cause decoherence in the
dynamics of a qubit. If we cannot follow the coupled dynamics of the
qubit and each individual bath spin, we must average over the
different possible time evolutions of them - a definite time
dynamics then gets converted to a time-evolving probability
distribution. Thus this entanglement between qubit and spin bath
dynamics causes decoherence in the qubit dynamics. In fact we can
classify three different decoherence mechanisms here, as follows:

(i) \underline{Precessional Decoherence}: The precessional motion of
$\vec{\sigma}_k$ around the time-varying $\vec{\gamma}_k$, shown in
Fig~\ref{precessF}, gives the strongest decoherence from a spin
bath.  Essentially the spin bath is absorbing \underline{phase} from
the qubit, and it turns out that the rate at which the qubit phase
dynamics is 'smeared' is roughly proportional to the average solid
angle swept out by all of the $\sigma_k$ together in their
precessional motion. A detailed calculation\cite{PS00,TupS04} gives
simple results for the decoherence rate $\Gamma_\phi^\rho$ in two
limiting cases:
\begin{eqnarray}
\Gamma_\phi^P &\sim& 1/2\sum_k(\omega_k/h_k)^2 ; \;\;\;\;\;\;\; (\omega_k\ll h_k )\\
\Gamma_\phi^P &\sim& 1/2\sum_k(h_k/\omega_k)^2 ; \;\;\;\;\;\;\;
(h_k\ll\omega_k)
 \label{gammaphiP}
\end{eqnarray}
where $\omega_k=|\vec{\omega_k}|$ in (\ref{ksigma}), and
$h_k=|\vec{h_k}|$. Note that this "precessional decoherence"
(originally called "orthogonality blocking"\cite{PSQTM94})
\underline{does not involve dissipation} --no energy is exchanged
between qubit and bath, only precessional phase. Thus this
decoherence mechanism, by far the most important one coming from the
spin bath, is not even dissipative!

The simplest and best-studied example of precessional decoherence is
the case of a spin qubit coupled to nuclear spins. Consider the
example of Fe$_8$, already discussed above, where the main
contribution to $\vec{h}_k$ is just the Zeeman coupling to the
external field, i.e., $\vec{h}_k\sim g_N\mu_N I_k H_0$, where $g_N$
and $\mu_N$ are the nuclear g-factor and magnetic moment. Then from
(\ref{gammaphiP}) we see that when $H_0=0$, the decoherence rate
$\Gamma_\phi=0$ (this is because $\vec{\gamma_k}$ flips then through
180$^\circ$; such a flip has no effect on the dynamics of
$\vec{\sigma}_k$). Decoherence then rises rapidly until reaching a
maximum when $h_k\simeq\omega_k$, where $\omega_k$ is the hyperfine
coupling as before - at this point the bath spins are precessing
wildly as the field $\vec{\gamma_k}$ flips between orientations
separated by angles $\sim 90^o$. Further increase in $H_0$ then
\underline{decreases} $\Gamma_\phi$, and in the high-field limit
$\Gamma_\phi$ again becomes very small (in this case
$\vec{\gamma}_k^\uparrow$ and $\vec{\gamma}_k^\downarrow$ are almost
parallel, so their flipping of $\tau_z$ again hardly affects the
dynamics). The quantitative details for Fe$_8$ were worked out in
ref.\cite{TupS04}.

(ii) \underline{Topological Decoherence}: The above discussion
concerned the bath spin dynamics when the qubit was sitting quietly
in either $\vert \uparrow \rangle$ or $\vert \downarrow \rangle$.
But what about when the qubit is actually flipping, and
$\vec{\gamma}_k(t)$ is moving rapidly between
$\vec{\gamma}_k^{\downarrow}$ and $\vec{\gamma}_k^{\uparrow}$? If
this flip were instantaneous, there would be no effect on the bath
spins - they simply couldn't follow the sudden change. However the
time $1/\Omega_o$ taken for $\vec{\gamma}_k$ to flip from
$\vec{\gamma}_k^\uparrow$ to $\vec{\gamma}_k^\downarrow$ is actually
finite (though still very short). Then elementary time-dependent
perturbation theory tells us that the amplitude $\alpha_k$ for the
spin $\sigma_k$ to make an inelastic transition under this sudden
change, ie., during the qubit flip itself, is $\alpha_k \sim \pi
\omega_k/2\Omega_0$. By the same arguments as before this bath
transition must cause decoherence in the qubit dynamics -- the bath
is actually \underline{registering} the change in the qubit state
(it is in effect performing a measurement), and this transition
actually adds a random topological phase to the qubit
dynamics\cite{PS93}. The resulting "topological decoherence"
contribution $\Gamma_\phi^T$ to the decoherence rate is found to
be\cite{PS00,PS93}:
\begin{equation}
 \Gamma_\phi^T\sim 1/2 \sum_k |\alpha_k|^2
 \label{topoD}
\end{equation}
For most systems this will be very small, since
$\omega_k\ll\Omega_0$ as a rule. Thus, for Fe$_8$, $\Omega_0\sim 5$K
in low applied fields, whereas most of the $\omega_k<1$mK; even
after summing over c. 200 nuclear spins, one still finds a very
small contribution to decoherence, changing slowly with field as
$\Omega_0$ decreases\cite{TupS04}.

(iii) \underline{Noise Decoherence}: The best-known kind of
decoherence is that coming from some unknown noisy vector field
$\vec{\chi}(t)$ coupling to $\vec{\tau}$. Chemists are very familiar
with this in NMR and EPR - it leads to the usual "$T_1/T_2$
phenomenology", in which the noise causes incoherent relaxation of
the qubit energy over a time $T_1$ and the qubit phase over a
timescale $T_2$. A spin bath also gives rise to this 'noise
decoherence'; the weak interactions between the bath spins cause a
slowly fluctuating field to act on the qubit, causing $T_1$ and
$T_2$ relaxation in the standard way.

Notice an interesting consequence of this, which is seen very nicely
in experiments on SMMs. Since most of the spin bath dynamics is
being driven by the qubit itself (see again Fig~\ref{precessF}),
then when the qubit is frozen by an external longitudinal bias field
$\epsilon\gg\Delta$, the bath spins must also largely freeze --
without the time-varying field $\gamma(t)$ to drive them, only the
very weak $V_{kk'}$ are left to give the spin bath any independent
dynamics (via the small slowly-varying $\vec{\xi}_k(t)$). Thus we
expect $T_1$ and $T_2$ to drastically increase when the qubit is
frozen. This was seen in very pretty experiments by the Leiden group
on the SMM Mn$_{12}$; they observed roughly a hundredfold increase
in $T_1$ and $T_2$ when they pushed the tunneling Mn$_{12}$
molecules off resonance\cite{morello04}.

\underline{Further remarks}: An obvious questions which arises here
is the following: to what extent are all these decoherence results
already incorporated into the $T_1/T_2$ phenomenology used in
experimental chemistry and physics in the last few decades?

We note first that the common practise in the literature
(particularly the quantum information literature) of associating
decoherence with some "experimental noise" fluctuation source, and
then applying standard ideas such as the fluctuation-dissipation
theorem\cite{fluctD} to associate this noise with a dissipation, is
quite wrong for a spin bath. As noted above, the most important part
of spin bath decoherence, i.e., precessional decoherence, simply
involves precessional motion -- there is no transfer of energy
between qubit and bath spins, and therefore no dissipation. Thus
there is no classical analogue to decoherence from a spin bath, and
no connection between decoherence and
dissipation\cite{PS93,PSQTM94,PS00,decoherence}.

Now the "$T_1/T_2$" phenomenology in NMR is based on the classical
Bloch equations. It throws away all the internal phase information
in the nuclear spin bath dynamics (including relative phases between
spins, entanglement, etc), leaving only the self-consistent
description of a single nucleus in the \underline{classical}
fluctuating mean field from all the other spins (plus any
fluctuating fields coming from electron spins, fluctuating external
fields, motional narrowing, etc).

\begin{figure}
\includegraphics[width=0.45\textwidth]{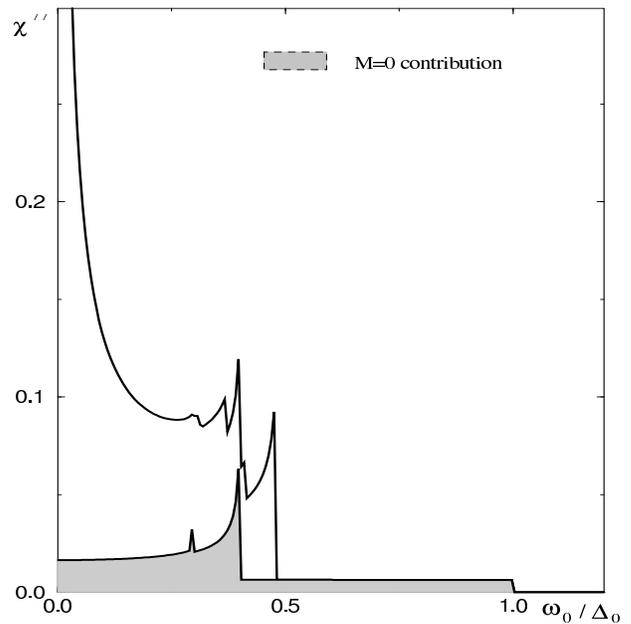}
 \caption{Plot of $Im ~\chi_{\uparrow\uparrow}(\omega)$, the
absorption probability described in the text, for a qubit coupled to
a spin bath. Without decoherence this would be a sharp line at
$\omega/\Delta_o = 1$. The shaded part shows contributions from
processes where the spin bath polarisation is unchanged.}
 \label{chiF}
\end{figure}

However this $T_1/T_2$ phenomenology, although it is commonly used
in the qubit literature (in discussions of, e.g., Rabi oscillation,
or spin echo for a qubit), it is not appropriate to the qubit
dynamics\cite{PS00,dube01}.  To give an idea of what can really
happen, in Fig.~\ref{chiF} we show the imaginary part of the Fourier
transform $\chi_{\uparrow\uparrow}(\omega)$ of the spin correlation
function $P_{\uparrow\uparrow}(t)$ for a qubit coupled to a spin
bath. This function is the probability that a qubit in state
$|\uparrow>$ at the time $t=0$ will be found there at time $t$. Thus
$\chi_{\uparrow\uparrow}(\omega)$ is the absorption intensity at
frequency $\omega$ in a qubit resonance experiment. This looks
nothing like the Lorentzian expected in the $T_1/T_2$ phenomenology,
and it shows singular peaks which come from the unusual coupled
dynamics of qubit and spin bath. These features actually result from
decoherence and are not associated with dissipation.

This point is important, because it means that eventually
experimentalists working on qubits are going to have to abandon the
$T_1/T_2$ phenomenology in describing their experiments, when spin
bath decoherence is important. The phenomenology that will be
required needs a whole paper on its own.

\subsection{Oscillator Bath Decoherence}

This has been well covered in other articles\cite{AJL87,grabert88}
and books\cite{weiss99}. The theoretical problem is simply
formulated -- one now couples a qubit to a bath of oscillators, and
tries to calculate the dynamics. This "spin-boson" model is well
understood\cite{weiss99}. In contrast to the spin bath, there is a
quantum-classical correspondence for oscillator
baths\cite{cal83,ajl84}, and a fluctuation-dissipation theorem
relating both the dissipative dynamics and the decoherence of the
quantum system to the fluctuational "quantum noise" acting on it,
coming from the oscillator bath.

Thus we can understand decoherence in the spin-boson model in a
fairly simple way. The oscillator bath has two main effects. First,
it slows down the qubit, increasing its "inertia", so that $\Delta$,
the qubit flip rate, is "renormalized" (ie.,
$\Delta\rightarrow\tilde{\Delta}$ with $\tilde{\Delta}<\Delta$).
Second, it causes dissipation in the "quantum relaxation" of the
qubit dynamics, either because the qubit spontaneously emits an
excitation (i.e., excites an oscillator from its ground state) or
scatters an existing excitation (i.e., stimulated emission -- an
excited oscillator changes its state, and possibly another one is
excited). All the complexity in the qubit dynamics comes from
repeated absorption and emission of oscillator
excitations\cite{feynman63,AJL87}. Nevertheless the phenomenology is
very familiar -- one defines a "spectral function"\cite{cal83} of
form
\begin{equation}
J_\alpha(\omega)=\pi/2\sum_q \frac{|c_q^\alpha|^2}{m_q
\omega_q}\delta(\omega-\omega_q)
 \end{equation}
(where $\alpha= \parallel,\perp$), and the entire dynamics can be
calculated in terms of this function, which is familiar from second
order perturbation theory or from Fermi's golden rule. It has the
form $(V^2/\omega)\times$(density of states), where $V\equiv
c_q^\alpha$ is a coupling constant and $\omega$ the energy
denominator, and the sum over $\delta(\omega-\omega_q)$ measures the
density of oscillator states at $\omega$.

Most of the theoretical work on the spin-boson system has been done
for the simple $T$-independent "Ohmic" form
$J(\omega,T)\rightarrow\eta\omega$. However most experiments,
certainly on qubit systems, are \underline{not} described by an
Ohmic model, for two reasons. First, as already discussed, spin bath
decoherence is not at all described by such a model. Second, even
without the spin bath, the oscillator bath is often not Ohmic. It
certainly will not be if the system is insulating, since for phonons
$J(\omega,T)$ depends strongly on T and is a higher than linear
power of $\omega$. Only for simple conductors is the Ohmic form
always valid. For a proper discussion of the different forms of
$J(\omega)$, see refs.\cite{AJL87,PS96,weiss99}.

As an example let us again take the SMM. The principal source of
oscillator bath decoherence is then the magnetoacoustic coupling to
acoustic phonons (for which $J(\omega) \sim \omega^3$. Both quantum
relaxation rates\cite{PS96,villain96} and decoherence
rates\cite{TupS04,MST06} have been calculated for this interaction.
The usual form for the decoherence rate in a transverse field is
\begin{equation}
\Gamma_\phi \sim \frac{|M_{f_i}|^2\Delta_o^3}{\pi \hbar^3 \rho
c_s^5} \coth(\frac{\Delta_o}{kT})
\end{equation}
where $M_{f_i}$ is the spin-phonon matrix element discussed in
section 2, $\rho$ the density, and $c_s$ the sound velocity. This
expression illustrates nicely the points made above; we have a
(matrix element)$^2$, multiplied by a density of states
$\sim\Delta_o^3$, divided by an energy denominator $\Delta_o$,
multiplied by a thermal factor $\sim\Delta_o^{-1} \coth
(\Delta_o/kT)$. In other words, just the golden rule. A more
sophisticated calculation hardly changes this result, simply
renormalizing $\Delta_o$. An Ohmic bath would give a completely
different result\cite{AJL87,PS96}. Actually, decoherence from an
Ohmic bath is typically very large, which is why in making a QIPS we
want to avoid having moving electrons (ie., one should use
insulating systems). Of course, this still leaves decohrence from
the spin bath.

\subsection{Pairwise Dipolar Decoherence}

We now turn to what may in the long run be the most insidious kind
of decoherence - that caused by unwanted long-range interactions
between the qubits. There are two kinds of these, viz. (i)
interactions like dipolar or (in the case of current-carrying
devices) unscreened inductive interactions, which can be treated as
instantaneous; and (ii) long-range interactions mediated by slow
particles like phonons, which may have a retarded character.

\begin{figure}
\includegraphics[width=0.45\textwidth]{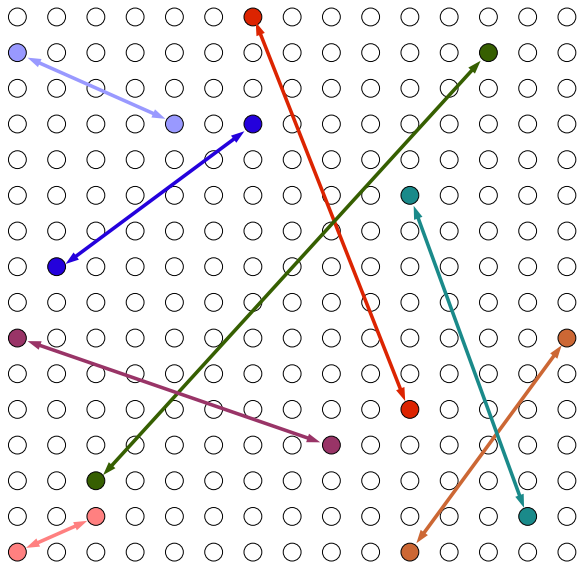}
\caption{2-dimensional spin network, showing resonant pairs of qubits
interacting via dipolar couplings.}
\label{dipdec}
\end{figure}

The reason such interactions are so dangerous is that they affect
many qubits at a time. Only a few theoretical studies have so far
appeared. Aharonov et al.\cite{AKP06} argue that concatenated error
correction codes can handle long-range interactions, but only just.
It is however likely that the model in this paper is
over-simplified, since detailed calculations for both insulating
systems\cite{MST06}(where dipolar inter-qubit interactions are
important) and conducting systems\cite{baranger} (where inter-qubit
interactions mediated by the electron bath are important) give
rather different results. Thus the problem is still open, and we do
not give any detailed results for decoherence rates here.

We can nevertheless give a simplified description of what is going
on (see Fig.~\ref{dipdec}). With long-range ($\sim 1/r^3$)
interactions, it is {\it always} possible for a given qubit to find
resonance with many others in 3 dimensions, whether we want it or
not, provided the strength of these interactions exceeds a critical
threshold (this result basically goes back to
Anderson\cite{PWAloc}). Typically for dipolar interactions the
resonances will occur over rather long distances. There are really
only two ways to get round this problem - first, to use error
correction (a topic whose details\cite{chuang} we have eschewed in
this article); and (ii) use a lower dimensional geometry, which
drastically reduces the number of qubits that are far apart
(compared to the 3d case). Ideally one uses both - in section 4 we
discuss certain architectures which can reduce the effective
dimensionality of the QIPS.

\section{Design criteria: possible architectures}

Let us now use the results given in the previous section to discuss
the best way to {\it make} a QIPS. To recapitulate - we want a
system of qubits whose mutual interactions, and the local fields
acting on them, can be accurately controlled (or, if they are static
in time, are very accurately fixed and can be tuned). We want to be
able to write quantum information onto the states of these qubits,
do a computation, and then read out the answer. We do not want stray
interactions or fields interfering with this, and we certainly want
to minimize decoherence caused by interactions with other
uncontrolled degrees of freedom. So how can we best achieve all
these desiderata?

As advertised in  the introduction, we concentrate here on designs
for spin QIPS using materials chemistry, ie., 'bottom-up'
approaches. The main advantages of such an approach (as opposed to
top-down 'nanofabrication' approaches), are:

A1) reproducibility - chemistry uses quantum mechanics to produce
absolutely identical molecules or other kinds of spin complex;

A2) Size - the qubits and other parts of the architecture are really
small, so that few degrees of freedom are involved, and few defects;
hence decoherence is reduced and timescales become short;

A3) No moving electrons - moving electrons always cause decoherence.
With insulating molecules or other spin complexes one can make a
true spintronics system, which only involves spin degrees of
freedom.

There are also disadvantages: the main ones are

D1) Size - molecular and ionic spin moments are currently too small
to be either controlled or examined individually, except in very
small numbers. In most QIP algorithms, we need to be able to
control, address, and read out the states of individual qubits.

D2) Dipolar interactions - as discussed above these are potentially
fatal to QIPS - one needs architectures that suppress them.

Our task is thus to find designs that use the advantages (A1)-(A3)
to their fullest extent, and overcome the disadvantages (D1) and
(D2) as much as possible. In bottom-up designs the main weapons we
have are (i) the methods of chemistry, and (ii) a huge variety of
potential geometries, ie., possible QIPS 'architectures'. We deal
with these in turn.

\subsection{Chemical approaches}

At the level of individual ions or molecules, materials chemists are
well-placed to deal with the problems of reproducibility and
decoherence, both at the nuclear and electronic scale.

\vspace{2mm}

{\bf (i) Nuclear Chemistry}: Unless one wishes to use nuclear spins
as a QIPS resource (for which see below), then their enormous
contribution to decoherence means we simply want to get rid of them.
Now while all "odd" nuclear isotopes (ie., with an odd number of
protons) have $I\neq0$, at least one isotope of every even element
has $I = 0$. Indeed, in natural mixtures of even-numbered elements,
the $I=0$ isotopes are the most common. Thus we should aim for
systems with even-numbered elements only.

Of course, if one aims strictly for even nuclei, the chemical
playground is dramatically reduced, and isotopic purification is
still typically required. Obvious transition metal choices are Fe
(with 2$\%$ $^{57}$Fe) or Ni (with 1$\%$ $^{59}$Ni). There are also
a number of rare earth candidates. These are either even-numbered
(like Nd, Dy, Er, Yb) with a complex mixture of isotopes and the
possibility of having only $I=0$ through costly isotopic
purification, or odd-numbered (like Tb, Ho), where $I\neq 0$ but
only a single isotope is naturally prevalent. Note however that in
this case we deal not with a single rare earth ionic doublet at low
energies, but an electronuclear pair of spins - the nuclear spin
hyperfine coupling is so strong that one cannot rotate without the
other\cite{moshe}.

An entirely new range of possibilities opens up if we use nuclear
spins as part of the QIPS (e.g., in a quantum register). We then
must be able to switch on and off the interaction (and hence the
information flow) between the nuclear and electronic spins. One way
to do this is dynamical - one tunes the electronic and nuclear spin
splittings (e.g., with external fields) so that resonant
hyperfine-mediated interactions can be switched on and off. Concrete
practical designs are needed here - the big stumbling block will be
decoherence from dipolar interactions between nuclear spins and
remote electronic spins (even if these latter do not partake in the
resonant interactions, their dipolar field adds extra dynamic phase
to the nuclear dynamics, and can throw them off resonance).

Another possibility is to switch the hyperfine interactions on and
off. In spin transition systems the electron spins can be switched
on and off without a change in the redox state, using temperature,
pressure or light (the LIESST effect,\cite{LIESST} or
HAXIESST\cite{HAXIESST} / SOXIESST\cite{SOXIESST} for hard / soft
X-rays). If a design based on this effect could be made to work it
would be very powerful - one would simply switch off the QIPS when
it was not needed, storing information in the nuclear register.

\vspace{2mm}

\begin{figure}
(a) \includegraphics[width=0.13\textwidth]{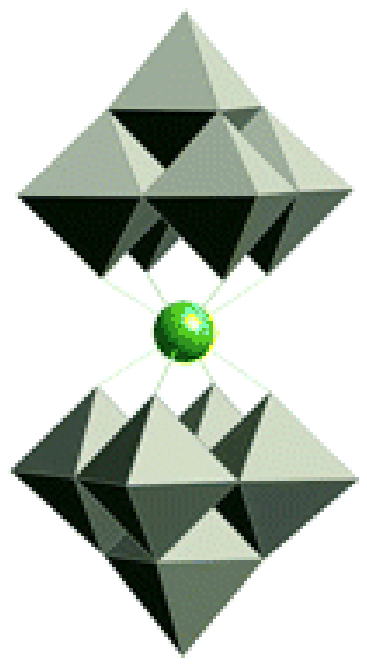} (b)
\includegraphics[width=0.26\textwidth]{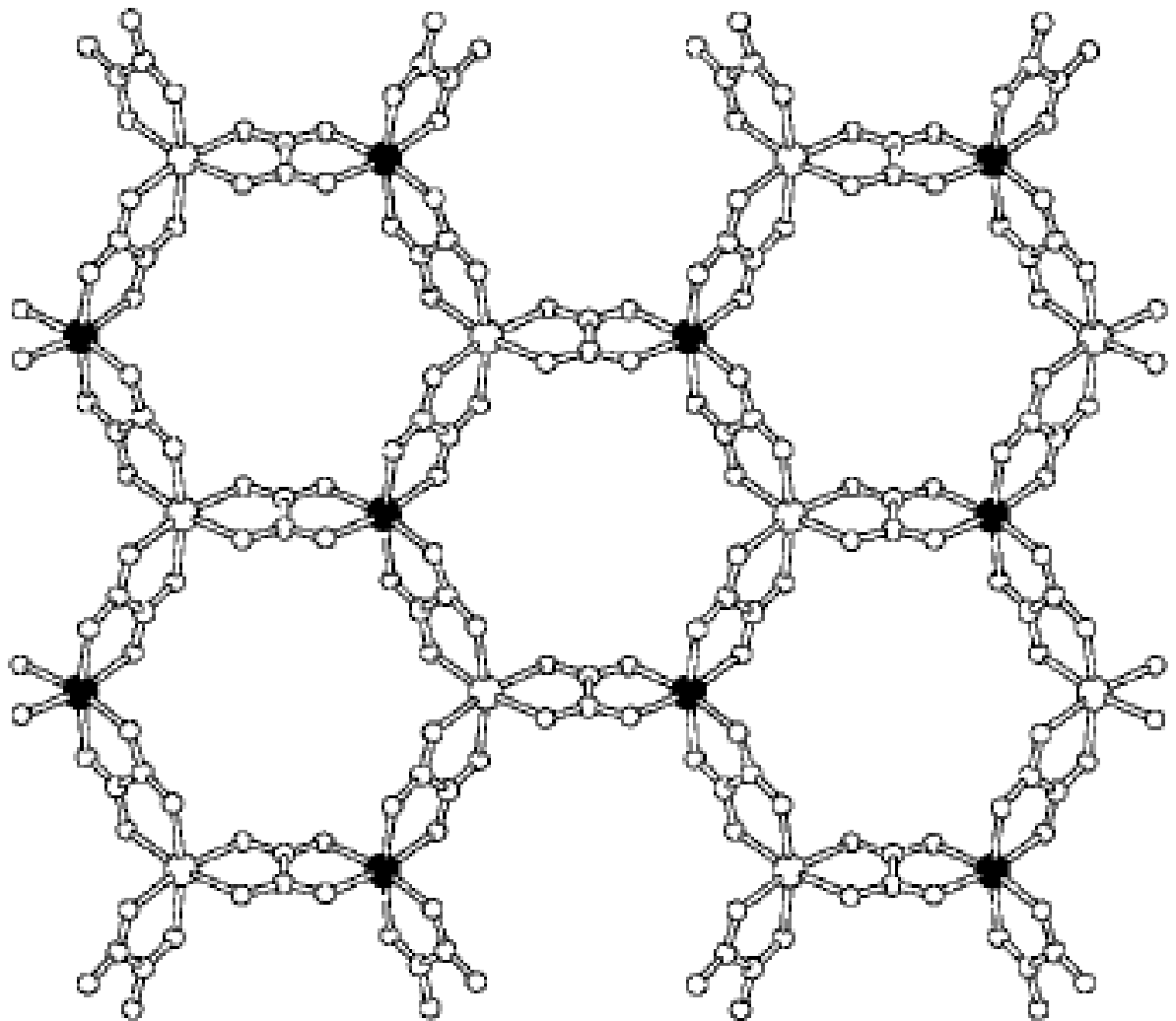} \caption{Some
systems that can be synthetized with a small amount of nuclear spins
and that could serve as hardware for qubits (see text for details):
(a) Single-molecule magnet based on a single lanthanide ion. (b)
Bidimensional "honeycomb" oxalate network.}
\label{chemsystems}
\end{figure}

{\bf (ii) Electronic Chemistry}: Undoubtedly one area where chemists
can play a big role is in the design of interactions (ie.,
engineering of effective Hamiltonians), and the preparation of
uniform \& reproducible qubits.

The first problem then is is the design of individual spin qubits.
Let us first note that seemingly obvious candidates such as
molecular Kondo systems or electrically-controlled quantum
dots\cite{QDot,molKondo,Mo12V2}, while promising, have one very
serious drawback - they have an intrinsically high decoherence, as
their operation is based on moving electrons.

So let us now consider insulating SMMs and rare earth ions as qubit
candidates. A question often asked by molecular chemists is - what
kind of crystal field is needed for the spin qubit? As noted
earlier, designs in which there is a 'natural basis' for the qubit,
created by a 2-well potential, are usually good ones. Many tunneling
SMMs and rare-earth ion molecular systems now use this design - the
question at issue is how big to make the spin anisotropy energy
barrier $E_B$. One wishes to operate the QIPS at energy scales $\ll
kT_c$, where $kT_c \sim E_B/2\pi S$ is the crossover energy to the
qubit regime (above $T_c$, the spin dynamics proceeds by incoherent
thermal activation). However at the same time we want to suppress
decoherence from nuclear spins and phonons. This imposes competing
restrictions: to reduce nuclear spin decoherence one wants $\Delta_o
\gg \{ \omega_k \}$, where the $\{ \omega_k \}$ are the couplings to
the spin bath (nuclear spins in this case), but to reduce phonon
decoherence one wants $\Delta_o$ as small as possible compared to
typical phonon energies (whose scale is determined by the Debye
energy $\theta_D$). Thus there is a 'coherence window'\cite{TupS04},
defined by $\theta_D \gg \Delta_o \gg \{ \omega_k \}$, defining a
desirable size for $\Delta_o$. Noting that $\Delta_o$ is certainly
less than $kT_c$, we can put these various restrictions together as
a requirement that $\theta_D \gg E_B/2\pi S \gg kT, \Delta_o \gg \{
\omega_k \}$. Thus we would like to have a large magnetic anisotropy
and weak hyperfine couplings - and a system that can go to very low
temperatures. Certainly the most important requirement here is weak
hyperfine couplings. It also helps if $\theta_D$ is big and if the
spin-phonon couplings are weak (which will happen if the molecule
is, eg., nearly spherically symmetric).

If one uses transition ion SMMs based on even nuclei (so as to
eliminate nuclear spin decoherence), severe restrictions are placed
on the chemical bridging between the spins - one has to discard all
the halogens and almost every organic molecule because of their
hydrogen and nitrogen nuclear spins. For coordination and
short-range bridging, obvious choices are then oxo, tio or oxalate
anions, but there are others, like carbonyl, thiocarbonyl or
carbide.

Alternatively one can go for rare earth-based molecules. Single
lanthanide cations in adequate coordination environments have shown
SMM behaviour\cite{Ishikawa}; and recently, polyoxometalates (POMs)
composed completely of W, O and Si have shown crystal fields capable
of giving qubit dynamics\cite{LnPOMSMM}. This offers exciting new
possibilities, as POMs are chemically quite different from the usual
SMMs (they have high negative charge, and are stable against
oxidation). Two advantages of any rare earth design are (i) the
relevant magnetic parameters can be calculated theoretically from
their structures and/or extracted from EPR and optical data; and
(ii) their strong magnetic signals and hyperfine coupling (when
using nuclei with $I>0$) facilitate their detection and
characterization.

Another key requirement is the reduction of disorder to very low
levels - this means eliminating defects, dislocations, etc., from
the QIPS, and using molecules or other spin complexes that only come
in one species (ie., not using systems that have active structural
degrees of freedom, various kinds of isomerism, or free rotation
about single bonds, or solvents/counterions that can take different
positions which are close in energy to each other).

Let us turn now to the design of inter-qubit interactions. These
need to be precisely controlled (whether they are static or dynamic)
and inter-qubit dipolar interactions and spin-phonon interactions
need to be suppressed. Since both are $\propto$ the qubit spin
moment, one possibility is to have antiferromagnetically ordered
spin qubits. The inter-qubit interactions would then vanish much
more quickly at long range (for an antiferromagnetic pair, they are
$\propto a/r^4$, where $a$ is the separation between the spins). If
the spins in each pair are not equivalent (eg., in an
antiferromagnetic transition metal dimer where the two metals are
either different or have different chemical environments), an
external stimulus (e.g. an EPR pulse) will differentiate the
$|\uparrow \downarrow>$ and $|\downarrow \uparrow>$ configurations.
This way of suppressing dipolar interactions also suppresses the
spin-phonon interaction. Important work that has already been done
in this direction with dimeric molecular
nanomagnets\cite{Mn4dim,Fe9,Mn12wheel}, including molecules where
there is a built-in switch\cite{Fe9}. We would suggest making
analogues to these systems in which the dipolar magnetic moments
cancel completely while a strong quadrupolar moment remains.
Alternatively, heterometallic antiferromagnetic
wheels\cite{antifwheels}, with $S=0$, are clearly promising building
blocks, and qubit designs for these systems have already been
explored\cite{antifwheels2}.

The most obvious way to implement inter-qubit interactions is to use
exchange or superexchange interactions between neighbouring spins.
Transition metal-based systems have interactions $\sim O(100K)$,
making them better than rare earths, where these interactions are
typically $\sim O(1K)$. If controllable long-range inter-qubit
interactions are desired, one can imagine propagating these through
reduced carbon nanotubes\cite{indexchnanotube} or
POMs\cite{indexchPOM}. Both can in principle mediate relatively
long-range indirect exchange through the delocalized electrons,
compared to superexchange, and both can be prepared using only $I=0$
isotopes. However, they involve moving electrons, a serious serious
source of decoherence; a better way would be to propagate spin
signals down a chain of strongly exchange-coupled insulating spins.

\subsection{Geometry, Architecture, and Fabrication}

The actual spatial arrangement of qubits, and the 'read in/read out'
probes which couple to them, is quite crucial to the design of a
QIPS. There are 3 main issues, viz. (i) the small size of spin
qubits based on molecules or other spin complexes means that with
current detection/control systems, it is hard to manipulate, read
in, or read out the state of a single qubit; (ii) the much larger
size of these detection/control systems means that they cannot be
'crowded in' to interact with more than a few qubits at a time; and
(iii) the geometrical design of a QIPS will greatly influence
decoherence, both from stray fields (e.g., from substrates) and from
long-range dipolar interactions.

\vspace{3mm}

{\bf (i) Geometry and Architecture}: Current probes (microSQUIDs,
STM and MFM systems, Hall probes, optical detection and control
systems, etc.) are partly dogged by the uncertainty principle - if
one is not careful, attempts to probe or control at a length scale
of order a qubit size (ie., a few nm) will have far too destructive
an effect on the qubits and their quantum states (thus if photons
were used, we would be dealing with soft $X$-rays!). STM and MFM
probes can have a very weak effect, even though their tips are very
small - but the rest of them is very big, and there is no obvious
way to address many qubits at a time with this technology.

The most obvious way to get round this problem is to use fixed
probes of {\it nm} size, connected remotely to the outside world
along one-dimensional connecters. Conventionally one thinks here of
wires and moving electrons - conceivably some future nanoSQUID array
could do the job. However it seems better to us to use a genuinely
spintronic arrangement of interacting spins (e.g., using
strongly-coupled chains of spins, or possibly a set of nanotubes) to
couple into the qubit array. These can themselves be connected to
much larger control and read in/read out systems outside the QIPS.
As yet there are no concrete designs in this area - it would be
useful to have some.

Another possibility, at least in the near future, is to make larger
spin-based qubits - a large (e.g., 50 $\times$ 50) spin array
(square, hexagonal, linear) of microscopic spins or SMMs, coupled
antiferromagnetically. These would interact via a few nearest
neighbour spins. A further advantage of this design would be
obtained if each qubit had no net spin, thereby suppressing dipolar
errors. Promising advances here are being made with the synthesis
and study of mesoscopic antiferromagnetic grids\cite{grid}.

Another question which we find interesting is the way in which both
the read in/read out and decoherence problems might be alleviated by
using a 'sparse architecture' for the qubits\cite{GRS08}. By this we
mean an arrangement of qubits on, e.g., a planar substrate, whose
total number does not increase linearly with the area of the system,
but more slowly. This allows easier access for probes, and also
strongly reduces errors caused by dipolar interactions. Many
geometries are possible here, ranging from coupled lines or nanorods
of various shapes to 'fractal' patterns on surfaces. The chemistry
and physical properties of the substrates will be important in
limiting these designs. For example, while
dendrimers\cite{dendrimers} could be a promising scalable support
for sparse architectures, steric hindrance usually induces disorder
in their structures, which would cause decoherence.

Finally, let us mention two other interesting alternative designs
which could alleviate the architectural problem. The first involves
using specific algorithms that avoid single-qubit addressing
altogether~\cite{Robert}; only the entire QIPS is addressed, but on
multiple occasions after a series of pulses. The other uses
molecular cellular automata~\cite{molcelaut} which only need
addressability at edges, thus reducing the addressability problem by
one dimension. Both of theses designs suffer because they require a
long time to carry out a computation - they can only work if
decoherence times are really long.

\vspace{3mm}

{\bf (ii) Fabrication}: This brings us back to the problem of
fabrication of these QIPS arrays. The range of possibilities is
enormous; we mention just a few. As far as SMMs are concerned, we
gave the main desiderata above - the interesting question is what
new possibilities would be useful to explore. One is POMs, which
have a rich chemistry offering different topologies, sizes, and the
ability to encapsulate magnetic metal ions in arbitrary ligand
fields and/or host delocalized electrons if desired\cite{POMrev}. As
noted above, they can be built entirely using even elements (mainly
W, Mo and O). Thus, one could explore "giant" POM
wheels\cite{giantPOM}, currently about 4nm in radius. It may be
possible to build a wheel which (i) is even bigger, (ii) is free
from nuclear spins (current designs contain many water molecules),
and (iii) has the magnetic network we want inside. Alternatively,
with an array of STMs, one could chemically prepare a self-assembled
monolayer of POMs or fullerenes, and inject a single electron on
each molecule directly under each tip. The chemistry would be quite
easy, and avoids nuclear spins, since fullerenes and POMs are based
on even-numbered elements. The main problems would be (i) electron
delocalisation in the molecules introduces new paths to decoherence,
and (ii) this scheme would shift part of the difficulty from the
chemistry to the nanoengineering.

We also note that a way to prepare antiferromagnetic 2-D lattices
which are poor in nuclear spins and have two distinct sites is to
use bimetallic honeycomb oxalate layers\cite{oxalatolayer}; see
Fig.~\ref{chemsystems}. One of the metals will always carry a
nuclear spin, but the other can be clean if one so chooses.  They
have no water, but do require a nearby cationic layer, to compensate
their own anionic charge. This cationic layer could be
Ca$^{2+}$-based.

Finally, we emphasize the care required in designing substrates.
They will cause strong decoherence if there are any 'spin bath'
defects (dangling bonds/free radicals, or charge defects, or
paramagnetic spins, or nuclear spins) which can couple to the
qubits. Solving this will not be easy - getting rid of nuclear spins
in the substrate may be the hardest task of all. POMs, among a vast
variety of molecules, have been organized bidimensionally in
different ways, including Langmuir-Blodgett films and through
covalent modification of the surface. Of course, usually these ways
make extensive use of elements with an odd number of protons, so
potentially the cleanest way would be to use self-assembled
monolayers (SAMs) on an adequate substrate. Highly ordered pyrolytic
graphite\cite{POMHOPG} and silicon\cite{POMSi} have been
demonstrated to support SAMs of - among other chemical species -
different POMs, and are all free from nuclear spins. Quartz would be
a good insulating substrate, but the preparation of POM SAMs on
quartz usually involves organic molecules to provide for a positive
charge\cite{POMquartz}, which defeats the purpose of choosing a
nuclear-spin-free substrate.

\section{Concluding Remarks}

The international effort to make a QIPS has assumed very large
proportions, but building a QIPS will obviously not be easy - the
main problem, as we have seen, is to suppress decoherence,
particularly from nuclear spins and dipolar interactions. The use of
insulating atomic or molecular scale spin qubits offers important
advantages - reduced decoherence, and with appropriate care, high
reproducibility. We have explored herein a large number of design
strategies for building a QIPS. It seems likely to us that
experimental efforts to gain control of decoherence and make systems
of multiply-entangled spin qubits will succeed in the next few
years. At this point it should be possible to devise realistic
architectures for large scale quantum information processing, and we
anticipate that chemical considerations will play a large role in
these efforts.

This work was supported by NSERC, CIFAR, and PITP in Canada, and by
MEC in Spain.



\end{document}